# Analyzing the Non-Functional Requirements in the Desharnais Dataset for Software Effort Estimation

Ali Bou Nassif and Luiz Fernando Capretz
Department of ECE, Western University
London, Ontario, Canada
{abounas, lcapretz}@uwo.ca

Danny Ho
NFA Estimation Inc.
Richmond Hill, Ontario, Canada
danny@nfa-estimation.com

*Abstract—* Studying the quality requirements (aka Non-Functional Requirements (NFR)) of a system is crucial in Requirements Engineering. Many software projects fail because of neglecting or failing to incorporate the NFR during the software life development cycle. This paper focuses on analyzing the importance of the quality requirements attributes in software effort estimation models based on the Desharnais dataset. The Desharnais dataset is a collection of eighty one software projects of twelve attributes developed by a Canadian software house. The analysis includes studying the influence of each of the quality requirements attributes, as well as the influence of all quality requirements attributes combined when calculating software effort using regression and Artificial Neural Network (ANN) models. The evaluation criteria used in this investigation include the Mean of the Magnitude of Relative Error (MMRE), the Prediction Level (PRED), Root Mean Squared Error (RMSE), Mean Error and the Coefficient of determination ($R^2$). Results show that the quality attribute "Language" is the most statistically significant when calculating software effort. Moreover, if all quality requirements attributes are eliminated in the training stage and software effort is predicted based on software size only, the value of the error (MMRE) is doubled.

## I. INTRODUCTION

Software requirements are mainly composed of functional requirements and non-functional requirements (NFR) [1]. Although the term NFR has been used for more than 25 years, there is still no agreement on the definition of NFR [2] and the term NFR is sometimes referred to "quality requirements". Kotonya and Sommerville define NFR as "requirements which are not specifically concerned with the functionality of a system." [3]. NFR play a pivotal role in the success or failure of software development [1]. Real-life software problems are more related to NFR than functional requirements as many disgruntled customers complain from systems that are characterized by poor quality, unreliable, inefficient, unsecure, etc. [1]. NFR are classified into three main categories [3]. These include product requirements, process requirements and external requirements. Examples of product requirements include software availability, safety, reliability and efficiency. Process requirements include requirements on the development standards such as the type of the programming language and CASE tools. External requirements are those that are derived from the environment and may be placed on both the product and process requirements.

In software effort estimation models, software effort is a function of software size and some other attributes such as project complexity, team experience, project type and language type, etc. [4]. Some of these attributes represent the quality requirements of the software. Although there is a unanimous agreement that quality requirements are very important and can be critical for the success of the project [2], it is not clear from the literature to what degree quality requirements are important in software effort prediction models. Some software effort prediction models such as the ones developed in [5], [6] and [7] completely ignore the quality requirements as software effort was predicted based on software size only. In other models [8], quality requirements attributes are represented by the technical factors of the project and can increase software effort up to 30%. Others [9] argue that quality requirements attributes represent more than 50% of the total effort.

This research investigates the importance of the quality requirements attributes for software effort estimation in the Desharnais dataset. The motivation of running this investigation on the Desharnais dataset is because this dataset has become very popular in addition to other datasets such as ISBSG [10] and COCOMO [11] for training and validating software cost estimation models.

In this paper, we ask three research questions related to the Desharnais dataset since these questions are not addressed in the literature:
1. What is the influence of each of the quality requirements attributes on software effort estimation?
2. What is the influence of all of the quality requirements attributes combined on software effort estimation?
3. Is any of the quality requirements attributes highly correlated to another quality attribute (multicollinearity)?

In this research, we show that some quality requirements attributes are statistically more significant than other attributes. Furthermore, ignoring the quality requirements attributes when calculating software effort causes an increase in the error from 31% to 62% based on the MMRE criterion during the training process using an ANN model.

The remainder of the paper is organized as follows: Section II provides an overview of the Desharnais dataset. Section III defines the terms used in the evaluation criteria. Section IV presents an investigation of the quality attributes using a multiple linear regression model as well as an Artificial Neural Network (ANN) model. Section V displays the results and provides general discussion about the results and in Section VI, we conclude the paper.

## II. DESHARNAIS DATASET

The Desharnais dataset [12] is composed of a total of 81 projects developed by a Canadian software house in 1989. Each project has twelve attributes which are described in table I. The projects 38, 44, 65 and 75 contain missing attributes, so only 77 complete projects are used.

## III. EVALUATION CRITERIA

In this paper, we used five different evaluation criteria. These include:

1. *MMRE:* This is a very common criterion used to evaluate software cost estimation models [13]. The Magnitude of Relative Error (MRE) for each observation i can be obtained as follows:

$$MRE_i = \frac{|Actual\ Effort_i - Predicted\ Effort_i|}{Actual\ Effort_i}. \quad (1)$$

MMRE can be achieved through the summation of MRE over $N$ observations:

$$MMRE = \frac{1}{N}\sum_{1}^{N} MRE_i \quad (2)$$

2. *PRED (x)* can be described as the average fraction of the MRE's off by no more than *x* as defined by [14]:

$$PRED(x) = \frac{1}{N}\sum_{i=1}^{N} \begin{cases} 1\ if\ MRE_i \leq x \\ 0\ otherwise \end{cases}. \quad (3)$$

The estimation accuracy is proportional to PRED (x) and inversely proportional to MMRE. PRED (0.25) is used as one of the evaluation criteria.

3. Root Mean Squared Error: The Root Mean Squared Error (RMSE) is the square root of the mean of the square of the differences between the actual and the predicted efforts as shown in Equation (4).

$$RMSE = \sqrt{\frac{\sum_{i=1}^{N}(E_{ai} - E_{pi})^2}{N}}. \quad (4)$$

where $E_a$ and $E_p$ are the actual and predicted efforts respectively, $N$ is the number of observations.

4. Mean Error:

$$\bar{x} = \frac{1}{N}\sum_{i=1}^{N} x_i. \quad (5)$$

Where $x_i = (E_{a\ i} - E_{pi})$

5. The coefficient of determination $R^2$: It is defined as the proportion of the variance in the dependent variable that is predictable from the independent variable. The value of $R^2$ varies between 0 and 1. An acceptable value of $R^2$ is $\geq 0.5$ [15].

TABLE I. PROJECT ATTRIBUTES

| Attribute | Variable Classification | Description |
|---|---|---|
| Project | Numeric | Project ID which starts by 1 and ends by 81 |
| TeamExp | Numeric | Team experience measured in years |
| ManagerExp | Numeric | Manager experience measured in years |
| YearEnd | Numeric | Year the project ended |
| Length | Numeric | Duration of the project in months |
| Effort | Numeric | Actual effort measured in person-hours |
| Transactions | Numeric | Number of the logical transactions in the system |
| Entities | Numeric | Number of the entities in the system |
| PointsNonAdjust | Numeric | Size of the project measured in unadjusted function points. This is calculated as Transactions plus Entities |
| Envergure | Numeric | Function point complexity adjustment factor. This is based on the General Systems Characteristics (GSC). The GSC has 14 attributes; each is rated on a six-point ordinal scale. $$Envergure = \sum_{i=1}^{14} GSC_i$$ |
| PointsAdjust | Numeric | Size of the project measured in adjusted function points. This is calculated as: $PointsAdjust = PointsNonAdjust \times (0.65 + 0.01 \times Envergure)$ |
| Language | Categorical | Type of language used in the project expressed as 1, 2 or 3. The value "1" corresponds to "Basic Cobol", where the value "2" corresponds to "Advanced Cobol" and the value "3" to 4GL language. |

IV. INVESTIGATION OF THE QUALITY REQUIREMENTS ATTRIBUTES

The goal of this research is not to evaluate which of the independent variables in the Desharnais dataset are statistically significant as this has been addressed in some studies [16], [17] and [18], even though each of these studies has a different output. The main goal is to study the influence of the quality requirements attributes on software effort estimation. For this purpose, we introduce a multiple linear regression model and an Artificial Neural Network (ANN) model.

*A. Multiple Linear Regression Model*

In multiple linear regression models, a dependent variable (aka target) is a function of two or more independent variables (aka predictors). In software effort estimation models, the dependent variable is "Software Effort" and the independent variables are those that are correlated to software effort. The main independent variable is "Software Size"; however, other independent variables are also important [19]. Based on Table I, there are twelve attributes. The first step before applying the regression model is to filter these attributes to see which is the dependent variable and which are the independent variables. Among these twelve attributes, there are two dependent variables (Length and Effort). Since only one dependent variable is required, the attribute "Length" is removed and the attribute "Effort" remains as the dependent variable. Regarding the other ten attributes, the attributes "Project" and "YearEnd" are removed because they are not correlated to software effort. Among the remaining eight attributes, there are two attributes that represent software size which are "PointsNonAdjust" and "PointsAdjust" and only one should be used as software size. PointsAdjust is calculated from PointsNonAdjust based on the formula presented in Table I. Some studies [16], [20] and [21] used the PointsAdjust and others [17] and [18] used the PointsNonAdjust. The difference between the two in the Desharnais dataset is not significant as the mean of the PointsAdjust is 282 whereas the mean of the PointsNonAdjust is 298 based on the 77 selected projects. In this research, PointsNonAdjust is used based on the recommendation of [22] and since the attribute "Envergure" is used as one of the independent variables. Based on this discussion, our analysis is based on the dependent variable "Effort" and seven independent variables which include "TeamExp", "ManagerExp", "Transactions", "Entities", "PointsNonAdjust", "Envergure" and "Language".

Based on the categories of the NFR presented in Section 1 and based on the description of attributes in Table I, we notice that among the seven independent variables, there are four project attributes that can be considered as quality requirements attributes which include "TeamExp", "ManagerExp", "Envergure" and "Language". For instance, the attribute "ManagerExp" becomes a quality requirement attribute if we say for example, "The manager experience of the project should be more than five years". In the Desharnais dataset, the project attribute "ManagerExp" lists the experience of the manager in each project and part of this research is to see if the experience of the manager would have an influence on predicting software effort. The independent variables are all numeric except "Language" which is categorical. Categorical variables should be converted to numeric using dummy variables before running regression analysis [16]. In this paper, the attribute "Language" is converted to two dummy variables "L1" and "L2" based on definition proposed in Table II.

TABLE II. LANGUAGE ATTRIBUTE

| Language | Description | L1 | L2 |
|---|---|---|---|
| 1 | Basic Cobol | 1 | 0 |
| 2 | Advanced Cobol | 0 | 1 |
| 4 | 4GL | 0 | 0 |

Simple linear regression is applied if the data (Effort and Size) are normally distributed [23]. We applied a normality test on the attributes related to "Effort" and "Size" (Effort, PointsNonAdjust, Transactions and Entities) and we found that they are not normally distributed. To normalize data, logarithmic transformation (ln) was applied on "Effort", "PointsNonAdjust", "Transactions" and "Entities". After logarithmic transformation, the data became normally distributed.

Before applying regression analysis, Stepwise regression was used to indicate the independent variables that are statistically significant at the 95% confidence level. The Stepwise regression shows that only the independent variables "PointsNonAdjust", "Envergure" and "Language" (represented by the two dummy variables L1 and L2) are statistically significant at 95%. Since we are investigating the quality requirements attributes, the attributes "TeamExp" and "ManagerExp" were not eliminated since they are part of the quality requirements attributes. The attributes "Transactions" and "Entities" were excluded as the result of the Stepwise regression. Another good reason to exclude "Transactions" and "Entities" is because these attributes are correlated with the attribute "PointsNonAdjust" since "PointsNonAdjust" is computed by the summation of "Transactions" and "Entities". If "Transactions" and "Entities" are used with "PointsNonAdjust", multicollinearity will exist. Multicollinearity means that there is a correlation between one independent variable and other independent variables. If multicollinearity is present, several problems will arise. The greater the multicollinearity, the greater the standard errors will be. When high multicollinearity exists, confidence intervals for coefficients tend to be very wide and coefficients will have to be larger in order to be statistically significant [24].

Based on the above rules, the multiple linear regression equation applied is described as:

$$\ln(Effort) = 1.46 + 0.88 \times \ln(Size) + 1.41 \times L1 + 1.38 \times L2 \\ - 0.0471 \times TExp + 0.0623 \times MExp + 0.0204 \times Env \quad (6)$$

Where Effort is software effort in person-hours, Size is the PointsNonAdjust, L1 and L2 represent the Language, TExp is the team experience measured in years, MExp is the manager experience measured in years and Env is the attribute Envergure.

Based on the Analysis of Variance (ANOVA), the P value of the model (Equation 6) is 0.000 which means that the null hypothesis (all coefficients of independent variables are 0) will be rejected. Table III shows the P value of each of the independent variables as well as the Variance Infraction Factor (VIF).

TABLE III. ANOVA FOR INDEPENDENT VARIABLES

| Attribute | P value | VIF |
|---|---|---|
| ln(size) | 0.000 | 1.31 |
| L1 | 0.000 | 2.565 |
| L2 | 0.000 | 2.378 |
| TExp | 0.258 | 1.51 |
| MExp | 0.089 | 1.507 |
| Env | 0.000 | 1.515 |

Based on ANOVA, if the P value of an independent variables is less than or equal 0.05, then this independent variable is statistically significant at the 95% confidence level. The P values in Table III show that all independent variables are statistically significant at the 95% confidence level except TExp and MExp which coincides with the results of the Stepwise regression. Although TExp and MExp are not statistically significant at 95%, they are still correlated to software effort to a certain degree but eliminating these attributes will not deteriorate the accuracy of the model. Nevertheless, we decided not to eliminate these two attributes because they are part of the quality requirements in the Desharnais dataset.

The purpose of the column VIF is to test if multicollinearity exists. Signs of multicollinearity start to appear if the VIF of any independent variable is greater than 5 [24]. Based on the VIF values in Table III, we conclude that there are no signs of multicollinearity, and this answers the third research question raised in Section I.

The algorithm used to analyze the NFR based on the regression model is described as follows (The quality requirements attributes are "Language" represented by L1 and L2, "TExp", "MExp" and "Env"):

1. Define: set S contains four quality requirements attributes ($i_1$ to $i_4$), n=0
2. Begin: independent variables =size, elements in S
3. Generate a multiple linear regression: ln(Effort)=f(ln(Size), S)
4. Calculate the error based on the five evaluation criteria defined in Section III by comparing the actual effort of a project against the estimated effort (dependent variable Effort in Step 3)
5. If n=5, Goto Step 11
6. If n#0, return the NFR $i_n$ to set S
7. n=n+1
8. If n#5, eliminate the quality requirements attribute $i_n$ from the set S, then Goto Step 2
9. Eliminate all members in Set S
10. Goto Step 3
11. End

*B. Artificial Neural Netwrok Model*

To thoroughly investigate the influence of the quality requirements, an Artificial Neural Network (ANN) model has been used in addition to the regression model. The type of the ANN model is Multilayer Perceptron of one hidden layer. The initial number of inputs of the ANN is 6 (Size, Env, TExp, MExp, L1, and L2). The elimination process of the NFR attributes in the ANN model is the same as described in the above algorithm. Please note that "L1" and "L2" are considered as one quality attribute which is "Language". The parameters of the ANN model are depicted in Table IV.

TABLE IV. TRAINING PARAMETERS OF ANN MODEL

| Hidden layer Activation Function | Logistic |
|---|---|
| Output Layer Activation Function | Linear |
| Algorithm | Conjugate Gradient |
| Maximum Iterations | 10,000 |
| Convergence Tolerance | 1.0e-5 |
| Min. Improvement Delta | 1.0e-6 |
| Min. Gradient | 1.0e-6 |

During training, 20% of the training rows were held out to prevent overfitting. Moreover, the number of the hidden nodes varies based on the number of the inputs of the model.

V. RESULTS AND DISCUSSION

Table V presents the results obtained from the multiple linear regression model as well as the ANN model. This table shows the influence of each of the four quality requirements attributes, as well as the four attributes combined on software effort estimation. A discussion about the results follows the table. Please note that the accuracy of the model increases when the MMRE, RMSE and Mean values are lower, but the values of PRED and $R^2$ are higher.

TABLE V. EXPERIMENTAL RESULTS

| Independent Variables | Multiple Linear Regression Model | | | | | ANN Model | | | | |
|---|---|---|---|---|---|---|---|---|---|---|
| | MMRE | PRED(.25) | RMSE | MEAN | $R^2$ | MMRE | PRED(.25) | RMSE | MEAN | $R^2$ |
| Size, Env, Language, TExp, MExp | 0.32 | 46 | 2305 | 325 | 79.3 | 0.31 | 49 | 2350 | 434 | 81.1 |
| Size, Language, TExp, MExp | 0.32 | 45 | 2409 | 525 | 74.8 | 0.33 | 48 | 2320 | 275 | 76.2 |
| Size, Env, TExp, MExp | 0.57 | 40 | 3029 | 702 | 48.3 | 0.58 | 35 | 2951 | 505 | 48.6 |
| Size, Env, Language, MExp | 0.32 | 50 | 2301 | 352 | 78.9 | 0.33 | 48 | 2299 | 252 | 78.7 |
| Size, Env, Language, TExp | 0.32 | 46 | 2370 | 377 | 76.9 | 0.32 | 48 | 2377 | 229 | 79.2 |
| Size | 0.61 | 35 | 3122 | 791 | 42.4 | 0.62 | 40 | 3081 | 652 | 42.3 |

Based on the results in Table V, we do not see significant differences between the regression model results and the ANN model results, so our findings that are based on the regression model coincide with the ANN model. For simplicity, we will use the multiple linear regression model for evaluation.

When all quality requirements attributes are used with Size as independent variables, MMRE has the value 0.32. PRED(0.25) = 46, this means that the MMRE of 46% of the projects is less than or equal to 25%. The $R^2$ value is 79.3 which means that 79.3% of the variation in Effort can be explained by the independent variables.

When the Env arrtibute was eliminated, the accuracy of the model worsened based on all evaluation criteria except the MMRE which was the same. Nonetheless, the accuracy of the model is still good even without the attribute Env ($R^2$=74.8%).

When the Language attribute was eliminated, we notice that the accuracy of the model deteriorated based on the five evaluation criteria. The MMRE became 57% which is relatively high. More importantly, the $R^2$ value dropped to 48.3 which indicates that the precision of the model is low ($R^2$ is less than 50%).

Regarding the TExp attribute, we notice that there is no change in the MMRE. The precision of the model slightly worsened according to the Mean and $R^2$ values. However, surprisingly the accuracy of the model improved based on the PRED (improvement from 46 to 50) and RMSE (improvement from 2305 to 2301). We also conclude that the contribution of TExp to the model's precision is insignificant.

Regarding the MExp attribute, there is no change in the model's accuracy after eliminating this attribute based on the MMRE, and PRED criteria. However, the RMSE, Mean and $R^2$ values were slightly better before the elimination of this attribute. An unexpected result regarding the MExp attribute was detected based on Equation (6). The coefficient of MExp is positive (0.0623), which means that the experience of the project manager positively correlates to software effort. In other words, if there are two projects that have exactly the same software size, team experience, Language type and Envergure, the project with higher MExp (higher manager's experience) requires slightly more Effort and this is unexpected.

Based on the above discussion, we notice that the attribute "Language" has the most significant correlation to software effort and the model's accuracy deteriorated by 25% (based on MMRE, from 32% to 57%) when this attribute was eliminated. The reason behind that is because in 4GL languages, one can write a piece of code (e.g. 1,000 LOC) in 5 hours. If the same piece of code is required to be written using Basic Cobol, 15 hours would be needed. We notice from this example that the effort to write this piece of code using Basic Cobol is 3 times the effort required to write the same piece of code (same software size) if using a 4GL language. The second most significant attribute is Envergure which the model's accuracy deteriorated by 4% when this attribute was eliminated based on the $R^2$ criterion. The contribution of TExp and MExp is insignificant; however, we found that TExp negatively correlates to software effort but MExp positively correlates to Software Effort. This answers the first research question raised in Section I.

Lastly, when all quality requirements attributes were eliminated, the model's accuracy deteriorated based on the five evaluation criteria. Furthermore, the MMRE error increased by approximately 100% (from 32% to 61%) after the elimination of all quality requirements attributes and this answers the second research question.

VI. CONCLUSIONS

This research focused on the investigation of the quality requirements attributes in the Desharnais dataset. The Desharnais dataset is a collection of eighty one projects developed by a Canadian software house. The Desharnais dataset has become very popular as many developers use it in addition to other datasets to train and evaluate software estimation models. Among the twelve attributes in the Desharnais dataset, four quality requirements attributes were defined. These include "Language", "TeamExp", "ManagerExp" and "Envergure". The analysis of the NFR attributes was carried out based on a multiple linear regression model as well as an Artificial Neural Network (ANN) model. The evaluation conducted was based on five criteria which include MMRE, PRED(0.25), RMSE, Mean Error and $R^2$. The results obtained showed that the NFR attribute "Language" is the most statistically significant attribute and eliminating this attribute in training the regression model causes an increase of MMRE by 25%. The next statistically significant attribute was "Envergure", followed by "TeamExp" and "ManagerExp" which are less significant. The attribute "TeamExp" negatively correlates with software effort whereas "ManagerExp" positively correlates to software effort. Results also showed that ignoring the quality requirements attributes and developing prediction models based solely on software size as an independent variable leads to a 100% increase in the MMRE error.